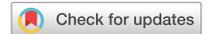

OPEN

# Identification of cohesive subgroups in a university hall of residence during the COVID-19 pandemic using a social network analysis approach

Pilar Marqués-Sánchez[1], Arrate Pinto-Carral[1✉], Tania Fernández-Villa[2], Ana Vázquez-Casares[3], Cristina Liébana-Presa[1] & José Alberto Benítez-Andrades[4]

The aims: (i) analyze connectivity between subgroups of university students, (ii) assess which bridges of relational contacts are essential for connecting or disconnecting subgroups and (iii) to explore the similarities between the attributes of the subgroup nodes in relation to the pandemic context. During the COVID-19 pandemic, young university students have experienced significant changes in their relationships, especially in the halls of residence. Previous research has shown the importance of relationship structure in contagion processes. However, there is a lack of studies in the university setting, where students live closely together. The case study methodology was applied to carry out a descriptive study. The participation consisted of 43 university students living in the same hall of residence. Social network analysis has been applied for data analysis. Factions and Girvan–Newman algorithms have been applied to detect the existing cohesive subgroups. The UCINET tool was used for the calculation of the SNA measure. A visualization of the global network will be carried out using Gephi software. After applying the Girvan–Newman and Factions, in both cases it was found that the best division into subgroups was the one that divided the network into 4 subgroups. There is high degree of cohesion within the subgroups and a low cohesion between them. The relationship between subgroup membership and gender was significant. The degree of COVID-19 infection is related to the degree of clustering between the students. College students form subgroups in their residence. Social network analysis facilitates an understanding of structural behavior during the pandemic. The study provides evidence on the importance of gender, race and the building where they live in creating network structures that favor, or not, contagion during a pandemic.

The COVID-19 health emergency has highlighted **the importance of relationships**, pointing to the most basic problems in our daily lives, such as problems of young people having sex[1], increased stress on parents in the face of restrictive measures that require their children to stay longer at home[2], or even hiding or lying about COVID-19-related symptomatology in order to keep a job[3]. The health emergency has even been identified as having a negative impact on the mental health of the population, especially due to a lack of social support[4]. The lower degree of social support has led to an increased risk of depression and sleep disturbances in the population[5].

This pandemic-induced scenario also requires attention with regard to the young population as they are very socially active and demand relationships especially outside the home, with their friends. One of the most important stages for a young person is the **university stage**. This stage implies a change in relationships and coexistence, an aspect of special interest in this pandemic context. In this sense, previous studies have already demonstrated the great psychological impact that the COVID-19 pandemic has had on university students[6].

[1]SALBIS Research Group, Department of Nursing and Physiotherapy, Universidad de León, Campus de Ponferrada s/n, 24400 Ponferrada, Spain. [2]The Research Group in Gen-Environment and Health Interactions (GIIGAS), Institute of Biomedicine (IBIOMED), Universidad de León, 24071 León, Spain. [3]Department of Nursing and Physiotherapy, Universidad de León, Campus de Vegazana s/n, 24071 León, Spain. [4]SALBIS Research Group, Department of Electric, Systems and Automatics Engineering, Universidad de León, Campus of Vegazana s/n, 24071 León, Spain. ✉email: apinc@unileon.es

Clearly, universities have had to implement numerous anti-pandemic measures, and social distancing is one of the most difficult measures to practice on university campuses[7]. In particular, student coexistence protocols in university residences is one of the most complicated aspects for academic authorities. These places are the closest hubs for students in terms of relationships. Students develop collective experiences that facilitate influences on each other, such as eating, sleeping or physical activity habits[8].

Therefore, knowledge of how their relationships are conducted during the COVID-19 pandemic in close living arrangements could be useful for hall-of-residence managers and university administrators. We recognise that it would be of interest to carry out interventions based on social structures in university residences in order to prevent contagion. But this requires a description of the actual scenarios, and there is still a lack of studies into this subject resulting from the rapidity of the pandemic process. The results of our research could be extrapolated to contexts of coexistence in which relationships are close: camps, schools, summer courses, extracurricular activities, etc.

In order to carry out an in-depth study based on the social structure of relationships, two decisions have been made. The first is to carry out a case study. The case study is useful in creating a conceptual framework about a real experience and use it as an illustration, therefore very useful in future research[9]. The second decision was to adopt social network analysis (SNA) as a methodological research approach to analyze the social structure of university students' relationships.

SNA is based on the relationships between interacting units in a network[10]. The elements of a social network are the nodes and links that represent an interaction between pairs of nodes, forming the social structure or network[11]. For Borgatti et al. structural issues are relevant[12]. He points out that in a team it is not only the abilities of its members to achieve a goal that matter, but the pattern of relationships that exists between them that is important. Therefore, the future characteristics of a node may depend on its position in the network structure[12]. The links are interconnected through shared nodes by drawing paths which, in turn, allow indirect connections to nodes that are not directly connected[13].

The application of SNA has been carried out in numerous health-related studies, e.g. SNA and physical activity[14], SNA and engagement in university rooms[15] and SNA and contagion in substance use[16], or intention to use cannabis[17] among others. There are also a significant number of SNA studies applied to the pandemic. For example, SNA studies on twitter conducted sentiment analyzes demonstrating the positive emotions of the population in coping with the pandemic crisis[18], studies into how social network structure can shape predictions on the spread and efficacy of control strategies[19], or the lack of communication between central and local government structures may have led to a lack of a unanimous response[20].

This network perspective, and the possibilities of transferring through connections, is particularly interesting for assessing which types of student social structures perform better or worse with regard to the COVID-19 contagion. We propose that contagion by COVID-19 in a space of close social relations is similar to the contagion of ideas, behaviors and customs. That is, the contagion of a pandemic, that is so dependent on social interaction, could be explained in the same way as **social contagion.**

Burt[21] points out that in a social structure there is social contagion in which a person becomes infected with a behavior, and this would occur among people who are perceived as socially similar and who are in groups with strong relationships. In order to carry out the study of cohesion among university students, we must assess the formation of subgroups. The assessment of subgroups will allow us to analyze the presence or absence of bonds and relate them to contagion by COVID-19. The presence of bonds could be interpreted as forces of social attraction to explain the creation of positive bonds within the group, an aspect that would reinforce group cohesion[22].

Social and community connectivity, moreover, is closely related to studies of community mobility and the spread of COVID-19. The more mobile the population, the more contact with the community and thus the greater the spread of the virus. Chang et al.[23] showed that reduced mobility helped to prevent infection. James et al.[24] also demonstrated the importance of analysing the mobility of people during the pandemic at the financial level. Finally, Yilmazkuday[25] corroborated the importance of analysing social and community connectivity through a mobility study using data obtained from Google, which showed a direct relationship between mobility and COVID-19 deaths over several weeks.

Social cohesion describes a series of events that facilitate interrelationships and can become a "force field" that conditions attitudes and behaviors among members[26]. Cohesive subgroups behave almost as independent entities, and this implies that there could be pressure from the group on its members to respect the norms or not[27]. Perceived norms become guidelines for action and, people decide which group to be in based on them. For this reason, it is really interesting in our research, since cohesive groups could have similar results in terms of positive or negative COVID-19 infections.

In the SNA there are several concepts for measuring subgroups. One of them is **cliques,** defined as complete subnetworks, within a network, of three or more nodes, in which all nodes are connected to each other[10]. However, this definition might be too strict for the object of this study without allowing us to evaluate the characteristics of the subgroups formed among the students in depth.

The question that really interests us is related to the connectivity characteristics of the subgroups, since it could provide us with information on the student networks in the university residences in relation to the COVID-19 infection. This structural information does not provide us with all the clues but it does indicate the most appropriate approach for the purpose of our research.

According to the above, our **research question** is: what are the structural characteristics of student subgroups in halls of residence in relation to the COVID-19 pandemic? To provide an answer, the following **aims** have been set out:

1. Analyze connectivity between subgroups based on aspects of the similarities between them.

| Scientific branches | Gender | | |
|---|---|---|---|
| | Female N (%) | Male N (%) | Total (%) |
| Engineering | 11 (44.0) | 2 (11.1) | 13 (30.2) |
| Health | 10 (40.0) | 11 (61.1) | 21 (48.8) |
| Social sciences | 4 (16.0) | 5 (27.8) | 9 (20.9) |
| Total (%) | 25 (58.1) | 18 (41.9) | 43 (100.0) |

**Table 1.** Sample characteristics.

2. Assess which bridges are essential for connecting or disconnecting subgroups[28].
3. To explore the similarities between the attributes of the subgroup nodes in relation to the pandemic context.

## Material and methods

**Study design.** This research was carried out using the case study methodology; developing a descriptive study. The case study facilitates decision-making on causalities and processes, provides answers to questions such as when and why, and is suitable for developing theory from the analysis of a process in contexts in which researchers do not yet have adequate answers[29]. The data was collected from 23th October 2020 to 20th November 2020.

**Setting and sample.** The participation was of 43 university students from different scientific branches but living in the same residence. The characteristics of the sample can be seen in Table 1. Of the total participants, 58.1% were men and 41.9% were women.

**Ethical consideration.** All participants received an informed consent form to participate in the study. Lastly, participants were offered the possibility of retracting consent once they had signed the form, without needing to provide a reason, and an email contact address was given should they require any further information. Participation was voluntary, and subject availability was respected at all times. All the participants that were involved in the study have given their informed consent to participate in this study.

This study has the approval of the Ethics Committee of the University of León under the identifier (ETICA-ULE-008-2021).

All data collected are related to the field of health. The relevant standards set out in Directive 03/2020 of the European Data Protection Committee have been complied with[30]. The data were anonymised and obtained after contacting the body responsible for COVID-19 contacts at the University of León, Epidemiological Surveillance System of the ULE (abbreviated in Spanish, SiVeULE). We confirm that all experiments were performed in accordance with relevant guidelines and regulations.

**Data collection.** SiVeULE is in charge of generating a database in order to follow up the COVID-19 cases that arise at the university. This database stores different characteristics of the actors in the network, together with the result of the RT-qPCR tests performed.

At the university there was a protocol to indicate norms and rules for (i) hygiene and preventive measures, (ii) what to do if you had symptoms, (iii) definitions of what was considered "close contact", "confinement" and "positive outcome". There were support staff to collect data, answer questions and help both positive and confined actors. These people were called "trackers". The name defined their role because they identified contacts of the student who were positive, had symptoms or had been "in close contact" with a positive person.

Other information such as name, residence, gender, grade, name of contacts, and date and result of the polymerase chain reaction (PCR) test are also collated in the database.

For the present study, the names were anonymized and registered in matrices for subsequent analysis using the SNA method.

The data obtained were used to construct a matrix that was read as follows:

- For rows, "A nominates B";
- For columns, "A is nominated by B".

To carry out this study, the matrix has been symmetrized, determining that if A nominated B, B also nominated A. That is to say, it is an undirected matrix, since, if A had any contact with B, B also had contact with A.

**Data analysis.** For data analysis, the social network analysis (SNA) has been applied to the $43 \times 43$ matrix. Factions and Girvan–Newman algorithms have been applied to detect the existing cohesive subgroups. These algorithms are frequently used for the detection of cohesive subgroups. For example, Vacca made use of Girvan–Newman to detect cohesive subgroups in his network[31], as did Zhu and Gao[32] in their study within a school classroom, results which they corroborated with feedback from teachers.

The factions algorithm is based on detecting subgroups made up of actors that result in obtaining densities as high as possible within the number of subgroups that are generated. That is, if we want to divide a network into 3 subgroups, it will create the subgroups that maintain a density as close to 1 as possible. If adding the three densities, the result is 3, we would have the optimal division of cohesive subgroups of the network[27].

This algorithm performs the following steps:

1. Arbitrarily assigning everyone to one of the groups and calculating their fitness to see if that partition is suitable.
2. It then moves some actors from one group to another and recalculates the fit again, observing if there is any improvement.
3. Continue in this way until you see no further improvement in the exchanges.

This method is known as "combinatorial optimizations".

In this case, using the factions algorithm, the network that we have used in this research was divided into different groups, 3, 4, 5, 6 and 7. After these divisions, the densities of the actors in each subgroup are analyzed among themselves and among the actors belonging to the rest of the subgroups. Thanks to this measure, and the resulting matrix, it is possible to determine which division generates subgroups with greater cohesion.

Because of the possible bias that may result from applying a single subgrouping algorithm, it was decided to use the Girvan–Newman algorithm. The Girvan–Newman algorithm also makes it possible to find out which division is the most optimal, by applying other different algorithms. In this case, Girvan–Newman looks for the most important links that, if removed, would fragment the network, reducing the cohesion of these subgroups. To apply this algorithm, the minimum number of cohesive groups to be generated is set at 3 and the maximum at 7. After applying this algorithm, a value, a metric, called "modularity", denoted by Q, is obtained. This measure compares the number of internal links in the groups and is higher as it finds groups that are more cohesive.

**Statistical analysis and visualisation.** IBM SPSS Statistics (26.0) software was used for the statistical processing of the data. For the analysis of descriptive data, frequencies and percentages were used for the qualitative variables, whereas the mean and standard deviation were used for the quantitative variables. A chi-squared test was carried out to verify whether there was a relationship between the groups. The UCINET tool, version 6.679[33] was used for the calculation of the SNA measurements. The tests carried out to study the normality of the distribution were the Shapiro–Wilk test because the sample is smaller than 55. The level of statistical significance was set at 0.05. For qualitative analysis, a **visualization** of the global network will be carried out using Gephi, version 0.9.2, software.

## Results

**Connectivity between subgroups.** After applying Girvan–Newman and Factions, in both cases it was found that the best division into subgroups was the one that divided the network into 4 subgroups. In the case of Girvan–Newmann, the test was carried out by dividing the network into 3 to 7 clusters, obtaining the best modularity with a division into 4 clusters, where Q = 0.439. By applying the factions algorithm, it was found that the best division was also made up of 4 clusters, obtaining a total density of 2.293, an average density of 0.573 for each group. In accordance with objective 1 presented in this manuscript, 4 cohesive subgroups have been obtained in which the actors are united because they are similar to each other.

**Bridges in subgroups.** As regards the assessment of which bridges are essential to connect or disconnect subgroups[28], in graphs 1 and 2 we can see that our entire sample of students is clearly divided into 4 subgroups (A, B, C and D). Both graphs show many edges within the subgroups resulting from the application of GN modularity, and few edges or bridges between them. One can therefore see extensive cohesion within the subgroups and low cohesion between them.

Figure 1 shows the subgroups in relation to positive or negative contagion. In general, it can be seen that there is greater cohesion between positive with positive and negative with negative. There are some key edges or bridges between the four clusters that could enhance contagion or non-contagion. This can be seen especially between the largest node that exists between cluster B, C and D. However, there are key bridges between infected and non-infected students that could be due to the living space. Figure 2 shows the students living together in one block of the dormitory in blue, and the students living together in another block of the residence in green.

**Attributes of the subgroup nodes and pandemic context.** *Group membership versus RT-qPCR result.* After obtaining the 4 cohesive groups, the relationship between group membership and RT-qPCR positivity was analyzed. The results obtained are set out in Table 2.

A chi-squared test of independence was carried out to examine the between subgroup membership and PCR+ percentage. The relationship between these variables was significant, $X^2 (3, N = 43) = 17.997$, $p < 0.001$. Subgroups A and B have a significantly lower percentage of PCR+ members than groups C and D.

*Group membership versus gender.* Subsequently, the relationship between group membership and gender was studied. The results are set out in Table 3.

A chi-squared test of independence was carried out to examine the between subgroup membership and gender. The relationship between these variables was significant, $X^2 (3, N = 43) = 13.480$, $p < 0.001$. Subgroups A and B have a significantly lower percentage of male and high percentage of female members while the opposite is true in groups C and D.

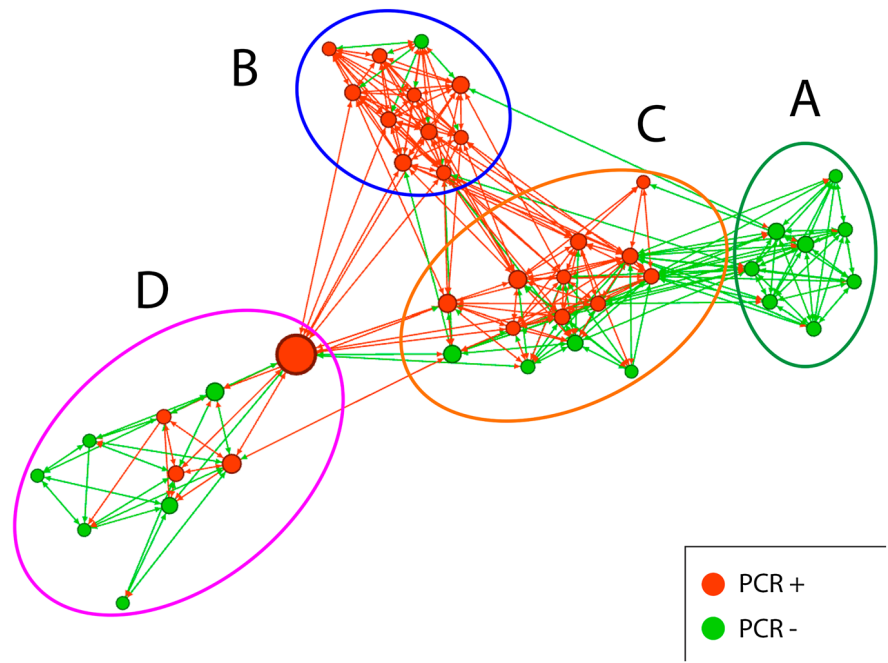

**Figure 1.** Subgroups Girvan–Newmann PCR result [created with Gephi 0.92 (https://gephi.org/)].

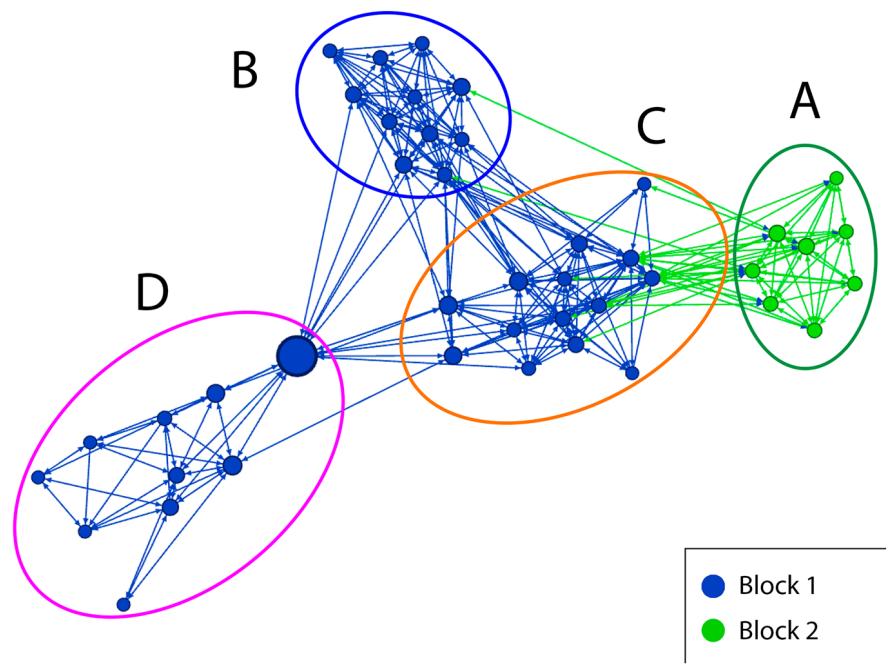

**Figure 2.** Subgroups Girvan–Newmann Residence block result [created with Gephi 0.92 (https://gephi.org/)].

*Group membership versus scientific branches.* Finally, the relationship between group membership and scientific branches of their members was studied. The results are set out in Table 4.

A chi-squared test of independence was carried out to examine the between subgroup membership and scientific branches. The relationship between these variables was significant, $X^2$ (6, N = 43) = 18.821, $p = 0.004$.

| Cohesive subgroups | PCR | | |
| --- | --- | --- | --- |
| | PCR+ N (%) | PCR− N (%) | Total (%) |
| A | 1 (9.1) | 10 (90.9) | 11 (25.6) |
| B | 4 (28.6) | 10 (71.4) | 14 (32.6) |
| C | 6 (60.0) | 4 (40.0) | 10 (23.3) |
| D | 8 (100.0) | 0 (0.0) | 8 (18.6) |
| Total (%) | 19 (44.2) | 24 (55.8) | 43 (100.0) |

**Table 2.** The relationship of actors belonging to the different cohesive subgroups and PCR results.

| Cohesive subgroups | Gender | | |
| --- | --- | --- | --- |
| | Female N (%) | Male N (%) | Total (%) |
| A | 5 (45.5) | 6 (54.5) | 11 (25.6) |
| B | 4 (28.6) | 10 (71.4) | 14 (32.6) |
| C | 8 (80.0) | 2 (20.0) | 10 (23.3) |
| D | 8 (100.0) | 0 (0.0) | 8 (18.6) |
| Total (%) | 19 (44.2) | 24 (55.8) | 43 (100.0) |

**Table 3.** The relationship of actors belonging to the different cohesive subgroups and their gender.

| Cohesive subgroups | Scientific branches | | | |
| --- | --- | --- | --- | --- |
| | Engineering N (%) | Social sciences N (%) | Health N (%) | Total (%) |
| A | 4 (36.4) | 6 (54.4) | 1 (9.1) | 11 (25.6) |
| B | 0 (0.0) | 6 (42.9) | 8 (57.1) | 14 (32.6) |
| C | 3 (30.0) | 2 (20.0) | 5 (50.0) | 10 (23.3) |
| D | 6 (75.0) | 1 (12.5) | 1 (12.5) | 8 (18.6) |
| Total (%) | 13 (30.3) | 15 (34.9) | 15 (34.9) | 43 (100.0) |

**Table 4.** The relationship of actors belonging to the different cohesive subgroups and scientific branches of their members.

In subgroups A and B there is a high percentage of social science students, while in groups C and D there is a higher percentage of engineering and health students.

### Discussion
Our research adds evidence to the scarce literature on COVID-19 with the application of social network analysis. Previous studies have looked at cohesion and groups in relation to sleep[34], and pandemic communication via twitter[35], and even how to deliver accurate messages to audiences of interest[36]. But our study would be the first to analyze cohesion in university students. Our findings analyze a population of special interest, as university students group together, share experiences and information, and this also takes place in a pandemic context.

We have outlined the potential of social networks to capture and understand the clustering among university students in their living arrangements. This research has focused on aspects of social cohesion in the context of the COVID-19 pandemic. Specifically, in the context of university residences, since close coexistence among university students could be key to both the spread and prevention of the virus. Previous studies have already found that social distancing and isolation are strategies introduced to combat pandemics[37]. However, given that university students have many relationships with many contacts and isolation does not seem feasible, it would be useful to study the structural characteristics of cohesion among students, and to assess how this relates to the degree of COVID-19 infection.

**Cohesive subgroups in the pandemic context.** Our results have shown that in university residences and in a pandemic context, clearly differentiated networks or reticular subgroups are formed. In order to understand this fact, it is necessary to understand the difference between "situational environment" of the student to the "reticular environment".

Lozares et al.[38] explain this clearly. The situational environment is based on each student's perception based on their daily experiences, such as, "I feel bored", "I am having fun", "I like the food", "this movie is fun", etc.

However, this perception in itself does not generate networks, since it is the reticular environment that represents the relationship between these everyday experiences. That is, the reticular or network environment means the relationship between events that occur in the students' everyday life. e.g. "I have meted my friends for eating" versus "I like eating", "I have met my friends to go to the cinema to see a fun movie" versus "I have fun", etc. With respect to this reticular context, a network is formed with both environments, the situational environment and the reticular environment. Therefore, in the network context, sub-clusters can be observed within the network, based on strong relationships in which actors are linked by shared events that generate a strong cohesion. This cohesion can become so strong that the performance of a grouping or sub-grouping can be similar to that of a single actor. It is therefore important to analyze cohesion on the basis of clusters.

Our findings show that infected students form subgroups, and uninfected students also tend to be more clustered. This could be related to the concept of social capital in networks. That is, social capital is the relational value of networks[39]. "I have more contact with the classmate who leaves me notes, or the one who listens to me, or the one who supports me when I am depressed. We must be attracted to someone, or someone must be attracted to us in order to group together.

Regarding contagion by COVID-19 and the importance of **how we perceive others**, Gould[40] highlights that contacts within small groups may be related to positive feelings among actors. For example, we prefer to receive positive feelings from contacts for whom we ourselves have positive feelings. In this sense, Homans[39] had already pointed out that feelings lead to interaction and vice versa. In view of this, students would not respond to pandemic protocols just because they know them. In this case, individuals are supposed to follow the rules for social responsibility. But knowledge does not guarantee action. It is the moral and emotional aspects that trigger certain social reactions that transform knowledge into action[41]. In this sense, it could be deduce that the moral and emotional aspects involved in the networks of coexistence among students could justify their behavior in the face of the pandemic and, consequently, their results on infection.

On the other hand, the actors also cohere according to their similarities. In this sense, our results show that gender could be a strong variable leading to the grouping of actors. This result would be in line with the concept of **homophily** and previous studies on the subject.

Homophily is the concept that people who are more similar contact each other than people who are dissimilar[42]. This means that cultural or behavioral aspects that are transferred through network channels will be localied and identified by the actors.

The concept of homophily is relevant in explaining clustering especially in a pandemic context. That is, local attachment between people who perceive themselves as more similar could bring about two phenomena. On the one hand local attachment, and on the other hand increased segregation from other groups[43]. This result could be positive or negative depending on the context of the study. For example, an increase in segregation between social networks in a neighborhood due to cultural diversity could lead to conflict and social division. However, if in a pandemic context, infected people are more clustered, the interpretation might be different. What will be more important if infected actors are more clustered or more dispersed? Given this question, it seems logical to think that it would be safer for the infected actors to be in the same bubbles. In this sense, it could also be considered more useful for non-infected students to be more dispersed, in order to be able to transmit responsible behavior. However, they themselves choose to group together, perhaps in an attempt to survive in the face of contagion. Thus, it seems that clustering in a pandemic context could be advantageous in order to address different strategies.

On the other hand, if clusters of infected students were very central, they could have put the rest of the clusters at risk. This is seen clearly in our study and the risk has also been explained in previous studies. Core clusters could become a "footprint" at the structural level of the network[44]. This fact could explain the possibility of liability or non-liability contagion behavior prior to COVID-19 protocols.

**Key bridges in the pandemic context.** Accessibility is important in the study of subgroups if the research raises the importance of bridges between groups or actors who can act as bridges in social processes[10].

Our results show that there are bridges between subgroups and that they are in line with previous studies that point to the importance of bridges between subgroups. This is because through them resources related to social capital such as information, support, trust and security are transferred and analyzed in healthcare contexts[45]. Occasionally, individuals may seek help or advice from outside their subgroup in order to achieve their objectives[46].

The ability to serve as a bridge in a network is given by the personal characteristics of the individual[47]. In this sense, we have collected some attributes that highlight the importance of gender and location within the university residence. In the network there are students with a position that clearly identifies them as bridges. This position could be related to these attributes, including the emotional attraction they arouse in the eyes of other students. Their capacity as intermediaries could serve as "glue" between the subgroups. Let us not forget that depending on the profile of the intermediary or bridge actor, the student could serve to spread behaviors that favor contagion or pandemic prevention. Previous studies have shown that if the aggregation of the groupings manifests feelings against the norms, it could increase contagions[48]. This confirms the importance of being aware of bridging links during a pandemic.

To conclude the discussion of our research, to underline that there is currently an excess of information about the pandemic, especially in digital networks. We might think that this information could influence students in university residences. But it seems that this excess of information could be confusing or even not credible for the younger population[49]. Therefore, their behaviors might be more influenced by what they think and how they interact with their close contacts than by online information. In this context, physical social interaction becomes relevant[50].

Obviously, this research has a number of **limitations.** One of the most important is the lack of further characteristics of the nodes with which to define the personality of the students in the network more accurately. Another significant limitation is the absence of more relational questions that could be useful in describing the case more accurately.

We recognise that the sample of the present research is small. However, the research team has considered that the case methodology would be useful for these contexts. Consequently, the results of this research would allow a first approximation of how networked individuals behave in a pandemic context. This information would be useful for future studies at the meso and macro level.

**For future research,** it is intended to replicate our study in more university residences even in the months at the beginning of the next academic year, since we will still live with preventive measures. It will be interesting to observe the behavior of the networks in this new situation.

### Conclusion

The present research has analyzed subgroups in a university hall of residence. The objectives addressed were the analysis of connectivity and bridging in the subgroups, as well as cluster node similarities in a COVID-19 pandemic context. The main conclusions are listed below:

- University students form networks in their residences. In turn, within the overall network, subgroups are formed.
- There are certain attributes that facilitate the groupings in this case study. Primarily these are gender and the block location in which the students live.
- The infection results are related to the subgroups, so that there are subgroups with a majority of infected students and other groups with a majority of uninfected students.
- Between the subgroups there are significant bridges, without which there could be a risk of disconnection between the clusters.
- There is a relationship between the scientific branches and the cohesive subgroups studied in this research.

### Acknowledgements
We are grateful for the work of all the students of our university who have helped during the COVID-19 pandemic to make the epidemiological surveillance system work as efficiently as possible.

### Author contributions
A.V.-C. and T.F.-V. conceived the project. J.A.B.-A., P.M.-S. and C.L-P. performed the analytical calculations. A.P.-C., and J.A.B.-A. performed all the numerical calculations. J.A.B.-A. and P.M.-S. wrote a first draft of the manuscript. All authors reviewed and edited the manuscript.

### Competing interests
The authors declare no competing interests.

### Additional information
**Correspondence** and requests for materials should be addressed to A.P.-C.

**Reprints and permissions information** is available at www.nature.com/reprints.

**Publisher's note** Springer Nature remains neutral with regard to jurisdictional claims in published maps and institutional affiliations.